\documentclass[aps,preprint,showpacs, endfloat]{revtex4}
\usepackage{graphicx}
\usepackage{dcolumn}
\usepackage{bm}
\begin{document}

\title{Giant Kerr nonlinearities and slow optical solitons in coupled double quantum-wells}

\author{Wen-Xing Yang}
\affiliation{Institute of Photonics Technologies, National Tsing-Hua University, Hsinchu 300, Taiwan}
\affiliation{Department of Physics, Southeast University, Nanjing 210096, China}
\author{Ray-Kuang Lee}
\affiliation{Institute of Photonics Technologies, National Tsing-Hua University, Hsinchu 300, Taiwan}
\email{rklee@ee.nthu.edu.tw}
\begin{abstract}
We show the formation of slow optical solitons in the asymmetric
coupled double quantum-wells (CQW) via a two-photon Raman resonance.
With the consideration of real parameters in AlGaAs-based CQW, we
indicate the possibility to have cancellation of the linear
absorption and giant Kerr nonlinearities.
With the controllable balance between dispersion and nonlinear effects in these solid-state based devices, this work may provide a practical platform for nonlinear optical signal processing.
\end{abstract}

\date{\today}
\maketitle
\makeatletter
\def\@dotsep{4.5}
\makeatother

Third-order Kerr nonlinearities play an important role in
nonlinear optics such as cross-phase modulation (XPM) for optical
shutters \cite{1} and generation of optical solitons \cite{2}, etc.
It is desirable to achieve giant Kerr nonlinearities with low light
powers \cite{3}. In recent years, both theoretically \cite{4} and
experimentally \cite{5}, the giant third-order nonlinear
susceptibility with reducing linear absorption has been one of the
most extensively studied phenomena. In addition, retaining the
merits of the giant Kerr nonlinearities, Wu and Deng \cite{6} have
theoretically proposed that it is possible to form ultraslow optical
bright and dark solitons for weak light by including the self-phase
modulation in cold atomic media.

However, it is more advantageous at least from the view point of
practical purposes to find, solid media that could permit to realize
the giant Kerr nonlinearities with low pump power, low absorptions,
and shape-invariant propagation of the optical field instead of the
aforementioned cold atom gases. In fact, we note that, in conduction
band of semiconductor quantum well (QW) structures, there have been
studies on the oscillations and wave propagations such as strong
electromagnetically induced transparency (EIT) \cite{7},
tunneling-induced transparency \cite{8}, ultrafast all-optical
switching\cite{9}, slow light propagation\cite{10}, etc. More
recently, the enhancement of Kerr nonlinearities based on
Fano-interference with intersubband transitions \cite{00} and a
large XPM have been studied in an asymmetric QWs \cite{11}.

In this Letter, we show that asymmetric semiconductor coupled double quantum-wells (CQW)
also can support the propagation of optical solitons via a
two-photon Raman resonance  scheme. Besides, under two-photon
resonance condition and with appropriate one-photon detuning, we can
obtain the cancellation of the linear absorption, enhancement of Kerr
nonlinearities, and slow group velocity propagation of the weak
probe pulse. Since the conduction subband energy level can be easily tuned by an external bias voltage, the proposed CQW structure also provide another possibility to realize electrically controlled phase modulator at low light levels.

Let us consider an asymmetric semiconductor CQW structure
consisting of 10 pairs of a 51-monolayer (145{\AA}) thick wide well
and a 35-monolayer (100{\AA}) thick narrow well, separated by a
Al$_{0.2}$Ga$_{0.8}$As buffer layer\cite{12}, as shown in Fig.
\ref{fig1}. The energy difference $2\delta$ of the bonding state
$\left| 3 \right\rangle$ and anti-bonding state $\left| 4
\right\rangle$ is determined by the level splitting in the absence
of tunneling and related tunneling matrix element, which can be controlled
by an electric field applied perpendicularly to CQW.

We assume the transitions $\left| 2 \right\rangle \leftrightarrow
\left| 4 \right\rangle $ and $\left| 2 \right\rangle \leftrightarrow
\left| 3 \right\rangle $ are simultaneously coupled by a strong
coupling field with the respective one-half Rabi frequencies $\Omega
_c = \mu _{42} E_c / 2\hbar $ and $(\Omega _c \mu _{32} ) / \mu
_{42} $. At the same time, a weak probe field is applied to the
transitions $\left| 1 \right\rangle \leftrightarrow \left| 4
\right\rangle $ with the respective Rabi frequencies $\Omega _p$.
And the transitions $\left| 1 \right\rangle \leftrightarrow \left| 3
\right\rangle$ is dipole forbidden transition due to selection
rules. $E_c $ and $E_p $ are the amplitude of the strong-coupling
 and weak probe field, respectively. By adopting the
standard approach \cite{13}, under the electro-dipole and
rotating-wave approximations the system dynamics can be described by
equations of motion for the probability amplitudes of the electronic
wave functions:

\begin{figure}[ht]
\centering
\includegraphics[width=4cm]{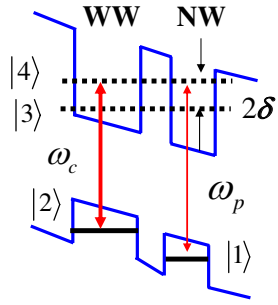}
\caption{Conduction subband energy level diagram for an asymmetric
coupled quantum wells consisting of a wide well (WW) and a narrow
well (NW).} \label{fig1}
\end{figure}

\begin{eqnarray}
\label{eq1} &&\hspace{-0.3in}\frac{\partial A_1 }{\partial t} = i\Omega _p^\ast A_4,\\
\label{eq2} &&\hspace{-0.3in}\frac{\partial A_2 }{\partial t} = - [\gamma _2 -
i(\Delta _p - \Delta _c )]A_2 + i\Omega _c^\ast A_4 + iq\Omega
_c^\ast A_3,\\
\label{eq3} &&\hspace{-0.3in}\frac{\partial A_3 }{\partial t} = - [\gamma _3 -
i(\Delta _p - \delta )]A_3 + iq\Omega _c A_2 + \kappa A_4,\\
\label{eq4}&&\hspace{-0.3in}\frac{\partial A_4 }{\partial t} = - (\gamma _4 - i\Delta
_p )A_4 + i\Omega _p A_1 + i\Omega _c A_2 + \kappa A_3,
\end{eqnarray}

\noindent where $A_{j} (j=1,2,3)$ being the amplitudes of subbands
$\left| j \right\rangle$. Here $\Omega_p = \mu _{41} E_p / (2\hbar)$
denotes one half Rabi frequencies for the transition $\left| 1
\right\rangle \leftrightarrow \left| 4 \right\rangle $, the
coefficient $q=\mu _{42} / \mu _{32}$ describes the ratio of a pair
of dipole moments with $\mu_{ij}$ being the dipole moment for the
corresponding transitions $\left|
i\right\rangle\leftrightarrow\left| j \right\rangle$. $2\delta = E_4
- E_3 $ is the energy splitting between the upper levels. $\Delta_c
= \omega _c - \omega _{42}$ and $\Delta_p = \omega _p - \omega
_{41}$ are the probe detunings of the coupling and probe fields with
transitions $\left| 2 \right\rangle \leftrightarrow \left| 4
\right\rangle$ and $\left| 1 \right\rangle \leftrightarrow \left| 4
\right\rangle$. The total decay rates $\gamma_{i}$ are given by
$\gamma_{i}=\gamma_{il}+\gamma^{dph}_{i}$, where $\gamma^{dph}_{i}$,
determined by intrasubband phonon scattering, electron-electron
scattering, and inhomogeneous broadening due to scattering on
interface roughness, is the dephasing decay rates. The population
decay rates $\gamma_{il}$, determined by longitudinal optical (LO)
phonon emission events at low temperature, can be calculated by
solving the effective mass Schr\"{o}dinger equation. For the
temperatures up to 10 K, the carrier density smaller than $10^{12}$
$\mbox{cm}^{-2}$, the dephasing decay rates $\gamma^{dph}_{i}$ can
be estimated according to Ref.\cite{12}.
$\kappa=\sqrt{\gamma_{3l}\gamma_{4l}}$ represents the cross-coupling
of states $\left| 3 \right\rangle$ and $\left| 4 \right\rangle$ via
the LO phonon decay. Note that a more complete theoretical treatment
taking into account these processes for the dephasing is though
interesting \cite{14} but beyond the scope of this work.

Under weak probe approximation ($\left( {\left| {\Omega _p } \right|
\ll \left| {\Omega _c } \right|} \right)^2$), almost all the
electrons are populated in the ground state $\left|1 \right\rangle$.
The excited state $\left| 4 \right\rangle$ can be adiabatically
eliminated when the variation of the probe field's envelope is slow
compared to the excited state lifetime, so there is no population
transfer of the ground state $\left|1 \right\rangle$. With these
assumptions, it is can be shown that $\left|{A_{1}}\right|^2\approx
1$, $A_{2,3,4}^{(0)}= 0$. With two-photon resonance condition
($\Delta_p=\Delta_c=\Delta$), we obtain the solutions of $A_j$ to
the first order of $\Omega_p$ from Eqs. (\ref{eq1}-\ref{eq4})

\begin{eqnarray}
\label{eq5} &&A_2^{(1)} = - \frac{(b - q\kappa )\Omega^{\ast}_{c}
\Omega _p }{abc - a\kappa ^2 + (b + cq^2 - 2q\kappa )\left| {\Omega
_c } \right|^2},\\
\label{eq6} &&A_3^{(1)} = - \frac{i(a\kappa + q\left| {\Omega _c }
\right|^2)\Omega _p }{ - abc + a\kappa ^2 - (b + cq^2 - 2q\kappa
)\left| {\Omega _c } \right|^2},\\
\label{eq7} &&A_4^{(1)} = - \frac{i(ab + q^2\left| {\Omega _c }
\right|^2)\Omega _p }{abc - a\kappa ^2 + (b + cq^2 - 2q\kappa
)\left| {\Omega _c } \right|^2},
\end{eqnarray}

\noindent with $a=-\gamma_2$, $b=-[\gamma_3-i(\Delta+\delta)]$, and
$c=-(\gamma_4-i\Delta)$. The first-order $\chi^{(1)}$ and
third-order $\chi^{(3)}$ susceptibility of the probe pulse are given
by \cite{6,0}

\begin{eqnarray}
\label{eq8} &&\hspace{-0.3in}\chi^{(1)}=-\frac{N\left| {\mu_{14}}
\right|^2}{\hbar\varepsilon_0}\frac{A_4^{(1)}A_1^{(1)*}}{\Omega_p},\\
\label{eq9} &&\hspace{-0.3in}\chi^{(3)}=-\frac{N\left| {\mu_{14}}
\right|^4}{3\hbar^3\varepsilon_0}\frac{A_{4}^{(1)}(\left|
{A_{4}^{(1)}} \right|^2+\left| {A_{3}^{(1)}} \right|^2+\left|
{A_{2}^{(1)}} \right|^2)}{\left| {\Omega_{p}} \right|^2\Omega_{p}},
\end{eqnarray}

\noindent where $N$ is the electron volume density. For the CQW
structure considered here, we take $\gamma_2=0$ for the lifetime of
level $\left| 2 \right\rangle$ by consulting Ref. \cite{15}. Based
on Eq. (\ref{eq8}), one can find that the first-order susceptibility
is mainly caused by cross coupling of the driving field $\Omega_c$.
When probe detuning $\Delta_p=0$, it shows that the absorption
Im$[\chi^{(1)}]$ and the dispersion Re$[\chi^{(1)}]$ of the probe
field do not depend on $\Omega_c$. We show in Fig. 2(a) their
dependences versus the probe detuning $\Delta_p$ with the same
parameter values used in Ref. \cite{12}. One can clearly see that
far away from the point of absorption peak, the linear absorption
will be closed to zero.

With two-photon Raman resonance  condition, the coupling of the
driving field with transition $\left| 2 \right\rangle
\leftrightarrow \left| 3 \right\rangle$ destroy the coherence
between state $\left|1\right\rangle$ and $\left|2\right\rangle$,
which causes the linear absorption of the probe field and also leads
to the nonlinear effect. As a result, $q\neq0$ indicates
constructive interference in XPM nonlinearities. We perform a
numerical calculation of the third-order nonlinear susceptibility in
Eq. (\ref{eq9}). As shown in Fig. 2(b), for a certain probe
detuning, for example, at the marker \textbf{B} in Fig. 2(b)
(corresponds to the marker \textbf{A} in Fig. 2(a) at the same
detuning frequency), linear absorption is vanished while the
strength of XPM is large, which suggests that large XPM can be
achieved with vanishing linear absorption. This interesting result
is produced by the cross coupling in the nonlinear susceptibility
associated with XPM. Unlike the cold atomic systems with specific
four-level atoms, the conduction subband energy varies with the bias
voltage. When we adjust the energy level of the bonding state
$\left| 3 \right\rangle$ and the antibonding state $\left| 4
\right\rangle$ at different bias voltages, different nonlinear phase
shifts can be obtained by such giant Kerr nonlinearity. Thus our
proposed CQW structures could be provided as a flexible device to
realize voltage control, solid-based phase modulators at low light
powers.

\begin{figure}
\centering
\includegraphics[width=4.2cm]{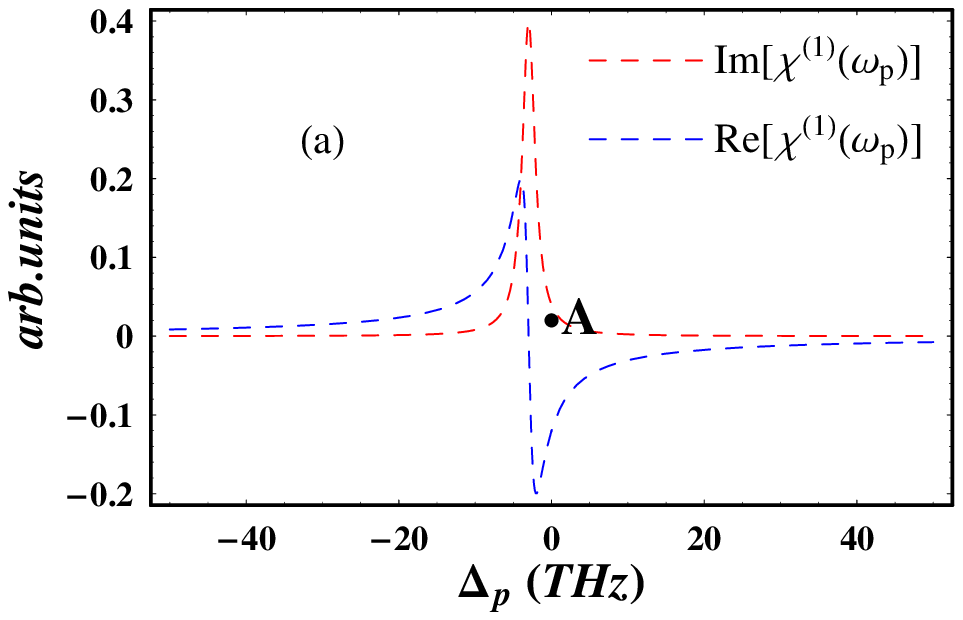}
\includegraphics[width=4.2cm]{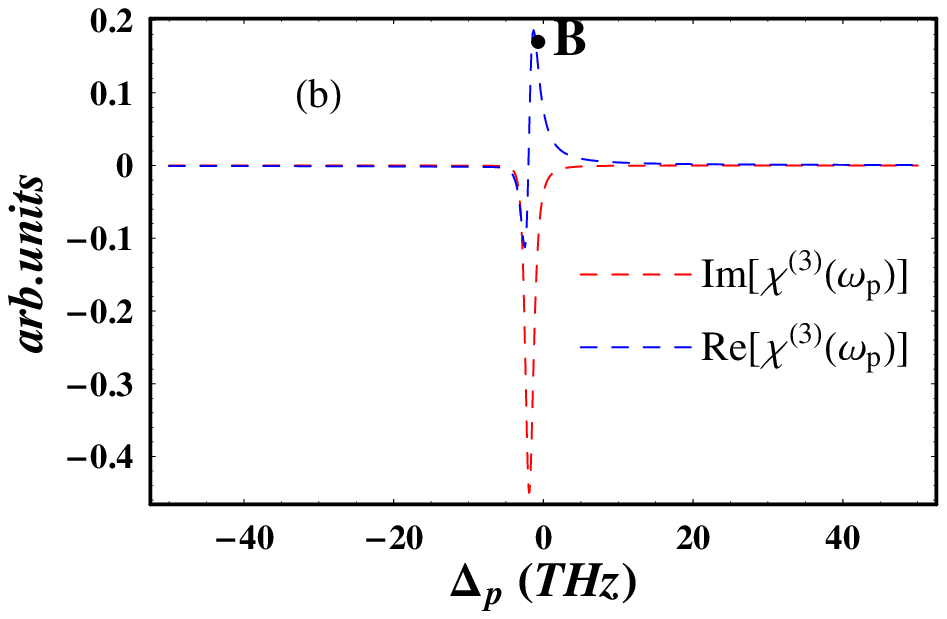}
\caption{ (a) Dependence of Im$[\chi^{(1)}]$ and Re$[\chi^{(1)}]$
and (b) dependence of Im$[\chi^{(3)}]$ and Re$[\chi^{(3)}]$ versus
the probe detuning $\Delta_p$. We have set $N\left| {\mu_{14}}
\right|^2/\hbar\varepsilon_0$ and $N\left| {\mu_{14}}
\right|^4/3\hbar^3\varepsilon_0$ as units in plotting, respectively.
The other parameters used are $\gamma_{3l}=0.7$ THz,
$\gamma_{4l}=0.9$ THz, $\gamma^{dph}_3=\gamma^{dph}_4=0.1$ THz,
$\delta=5$ THz, $\Omega_c=1$ THz, and $q=1.25$.}
\end{figure}

If the losses of the probe pulse are small enough to be neglected,
the balance between the nonlinear self-phase modulation and group
velocity dispersion (GVD) may keep a pulse with shape-invariant
propagation. From above discussion, as long as probe detuning
$\Delta_p$ is far away from the point corresponding to the
absorption peak, the linear absorption of the probe pulse is
negligible and the nonlinear self-phase modulation is enhanced. In
the slowly varying amplitude approximation, the wave equation of the
slowly varying envelope $E_p(z,t)$ of the probe pulse along $z$-axis
is given by \cite{0}

\begin{equation}
\label{eq10} (\frac{\partial}{\partial
z}+\frac{1}{v_{g}}\frac{\partial}{\partial t})E_p +
\frac{i\beta}{2}\frac{\partial^2}{\partial t^2}E_p =
\frac{2i\omega_p}{c}n\left| {E_p} \right|^2E_p,
\end{equation}

\noindent together with $v_g= c/Re[n_0+\omega_pdn_0/d\omega]$,
$n_0=\sqrt{1+4\pi\chi^{(1)}}$, $\beta=d^2k/d\omega^2$,
$n=3\pi\chi^{(3)}/n_0$, and $k=\omega_pn_0/c$ being the group
velocity, linear index of refraction, GVD, Kerr-nonlinear refractive
index, and wave vector, respectively, where $c$ is the light
velocity in vacuum. $v_g$ and $\beta$ are mainly determined by
Re$[d\chi^{(1)}/d\omega_p]$ and Re$[d^2\chi^{(1)}/d\omega^2_p]$,
respectively. We get the transformation of Eq. (\ref{eq10}) by
defining $\xi=z$ and $\tau=t-z/v_g$,

\begin{equation}
\label{eq11} \frac{\partial \Omega_p}{\partial
\xi}+\frac{i\beta}{2}\frac{\partial^2\Omega_p}{\partial\tau^2}=iW\left|
{\Omega_{p}} \right|^2\Omega_p,
\end{equation}

\noindent with $W=-2\pi\omega_p\chi^{(3)}/cn_0$, from Eq.(9). We can
choose reasonable and realistic set of parameters to satisfy
$\beta=\beta_r+i\beta_i\simeq \beta_r$ and $W = W_r + iW_i \simeq
W_r$, so that Eq. (\ref{eq11}) reduced to a standard nonlinear
Schr\"odinger equation which admits dark ($\beta_rW_r>0$) and bright
($\beta_rW_r<0$) solitons. The fundamental dark soliton takes the
form

\begin{equation}
\label{eq12}\Omega_p =
\Omega_{p0}\mbox{tanh}(\tau/\tau_0)\mbox{exp}(-i\beta_r\xi/2\tau^2_0).
\end{equation}

\noindent with $\left| {\Omega_{p0}\tau_0} \right|^2=-\beta_r/W_r$.
As an example, by taking $\Delta_p=10$ THz, we obtain
$v_g=0.9\times10^{-4}c$, $\left| {\Omega_{p0}\tau_0}
\right|=\sqrt{\beta_r/W_r}\simeq42.8$, and the linear absorption
coefficient $\alpha\simeq0.0066\mbox{cm}^{-1}$. There are four
adjustable parameters in our proposed system, i.e. the intensity of
the driving field, probe detuning $\Delta_p$, the energy splitting
$2\delta$ between the two upper levels, and the relative coupling
ratio$q$. From our numerical calculations, we find that decreasing
the Rabi frequency of the control field will decrease the group
velocity of the probe pulse and increase the values of $\beta_r$,
$W_r$ and $\left| {\Omega_{p0}\tau_0} \right|$. The control field
only need to be strong enough to couple two transitions $\left| 2
\right\rangle \leftrightarrow \left| 4 \right\rangle $ and $\left| 2
\right\rangle \leftrightarrow \left| 3 \right\rangle $, on the other
hand relatively lower intensity of driving field can lead to better
effects in formation of slow optical solitons. In addition, we have
used assumption of $\left| {\Omega_{p0}} \right|^2\ll\left|
{\Omega_{c}} \right|^2$ in our calculations, so the pulse width of
the probe field $\tau_0$ should be chosen to satisfy $\left|
{\Omega_{p0}\tau_0} \right|^2=-\beta_r/W_r\ll\left|
{\Omega_{c}\tau_0} \right|^2$. Fig. 2 shows that the parameter
$\Delta_p$ has a very large range of validity. For the parameters
$\delta$ and $q$, smaller separation $\delta$ will be better, which
is adjusted and there is no strict requirement for $q$.

It is worth noting that the cross coupling of control field may be
viewed as the perturbation to the two-photon resonance condition,
which comes from the closely separated two upper levels instead of
coming from the perturbation by introducing another laser field or
taking two-photon detuning \cite{6}, thus our scheme is a very
stable system to form slow optical solitons.

In conclusion, based on the two-photon Raman resonance scheme in
the asymmetric coupled double quantum-wells, we have shown that the quantum interference caused by cross
coupling of a strong CW laser field not only suppresses linear absorption loss, but also enhances Kerr nonlinearities of the weak probe pulse.
With the unique feature of controllable balance between linear dispersion and nonlinear effects in these solid-state devices, we also demonstrate the possibility to form ultraslow optical solitons.

The research is supported in part by NSFC under Grant Nos. 10704017
and 10575040, by NFRPC 2005CB724508. We would like to thank Ite. Yu
for his enlightening discussions.

\end{document}